\documentclass[aps,twocolumn,showlabels,showrefs,amsmath,amssymb,prl,superscriptaddress,floatfix,colors]{revtex4-1}

\usepackage{graphicx}
\usepackage{dcolumn}
\usepackage{bm}
\usepackage{amssymb}
\usepackage{hyperref}
\usepackage{multirow}
\usepackage{color}

\begin{document}

\title{Coexistence of defect morphologies in three dimensional active nematics}

\date{\today}

\author{Pasquale Digregorio}%
\email{lino.digregorio@ub.edu}
\affiliation{Departament de F\'{\i}sica de la Mat\`eria Condensada, Universitat de Barcelona, Carrer de Mart\'{\i} i Franqu\'es 1, 08028 Barcelona, Spain}
\affiliation{Universitat de Barcelona Institute of Complex Systems (UBICS), Universitat de Barcelona, 08028 Barcelona, Spain}
\author{Cecilia Rorai}%
\affiliation{CECAM, Centre Europ\'{e}en de Calcul Atomique et Mol\'{e}culaire, \'{E}cole Polytechnique F\'{e}d\'{e}rale de Lausanne (EPFL), Batochime, Avenue Forel 2, 1015 Lausanne, Switzerland}
\author{Ignacio Pagonabarraga}%
\email{ipagonabarraga@ub.edu}
\affiliation{Departament de F\'{\i}sica de la Mat\`eria Condensada, Universitat de Barcelona, Carrer de Mart\'{\i} i Franqu\'es 1, 08028 Barcelona, Spain}
\affiliation{Universitat de Barcelona Institute of Complex Systems (UBICS), Universitat de Barcelona, 08028 Barcelona, Spain}
\author{Federico Toschi}%
\affiliation{Department of Applied Physics and Science Education, Eindhoven University of Technology, Den Dolech 2, 5600 MB Eindhoven, Netherlands}
\affiliation{CNR-IAC, I-00185 Rome, Italy}

\begin{abstract}
We establish how active stress globally affects the morphology of disclination lines of a three dimensional active nematic liquid crystal under chaotic flow.
Thanks to a defect detection algorithm based on the local nematic orientation, we show that activity selects a crossover length scale in between the size of small defect loops and the one of long and tangled defect lines of fractal dimension $2$.
This length scale crossover is consistent with the scaling of the average separation between defects as a function of activity.
Moreover, on the basis of numerical simulation in a 3D periodic geometry, we show the presence of a network of regular defect loops, contractible onto the $3$-torus, always coexisting with wrapping defect lines.
While the length of regular defects scales linearly with the emerging active length scale, it verifies an inverse quadratic dependence for wrapping defects.
The shorter the active length scale, the more the defect lines wrap around the periodic boundaries, resulting in extremely long and buckled structures.
\end{abstract}

\maketitle

{\it Introduction}.
Local active stresses in a highly viscous nematic liquid crystal lead to the formation of sustained chaotic flow.
Such striking behavior has been observed in experiments of microtubules and kinesin molecular motors powered by ATP~\cite{sanchez_microtubules,guillamat_microtubules,lemma_2Dnem,martinez_actturb,tan_chaos}, as well as in bacterial suspensions~\cite{dombrowski_bacteriaturb,yeomans_livingturb}, epithelial cells~\cite{blanch_cellturb}, synthetic self-propelled particles~\cite{turbulence_in_janus}.
As internal energy is converted into a stirring at the scale of the single constituents, these systems are referred to as active nematics~\cite{marchetti_rev,gompper_roadmap}.

Activity induces out-of-equilibrium chaotic flow patterns at low Reynolds number, and this spatiotemporal state shares qualitative properties with inertial turbulence~\cite{toschi_2D}.
Many attempts have been made to identify a general framework and the universality classes for these phenomena.
In the context of active systems with nematic order, it has been shown that scaling regimes exist for the kinetic energy and enstrophy spectra, characterized by universal exponents~\cite{alert_actturbrev}.
However, differently than classically driven fluids at high Reynolds number, where a mechanism of energy cascade plays a key role in sustaining the turbulent behaviour, no energy transfers across scales have been observed so far for active chaotic flow~\cite{alert_scaling,urzay_turbulence,livio_cascade,livio_multiscale}.
The transition to a chaotic state in active nematics proceeds as an instability at the large scale of the nematic orientation, driven by the presence of local active stresses~\cite{edwards_act_nem}.
As the magnitude of the stress increases, smaller wavelength modes become unstable, generating spatiotemporal chaos.

Such transition generally leads to a proliferation of topological defects.
Contrary to passive systems where defects of opposite sign annihilate, in active nematics they are also steadily generated because of the energy input at small scales, and their non trivial dynamics is strongly correlated with the chaotic flow~\cite{thampi_turbulence,thampi_defects,giomi_defects,shankar_defects}.
The study of the statistics and dynamics of topological defects provides fundamental understanding of the phenomenology and the critical behaviour of systems at equilibrium~\cite{mermin_defects}, as well as active systems~\cite{nostro_difetti,marchetti_defects}.
Regarding chaotic flow driven by activity, numerous studies have focused on 2D active nematics, revealing that the key properties of the chaotic flow can be inferred from the behaviour of the $\pm1/2$ disclinations, as they are closely correlated with the fluid local vorticity~\cite{giomi_ActTurb}.
Recent results in 3D active nematics show that disclination lines are spontaneously generated in a regime of well-developed chaotic flow.
The primary defects excited in bulk active nematics are closed loops that undergo complex dynamics~\cite{toschi_exp}, as they are locally, and non-uniformly advected by the flow~\cite{ziga_defectcoarsening,houston_loops,toschi_exp,long_discl,binysh_loops,copar_droplet}.
Alongside with isolated loops, a complex network of disclination lines has been observed in experiments and simulations~\cite{toschi_exp,krajnik_spectra}.
The volume density occupied by defects increases with activity, and disclination lines continuously recombine  chaotically, providing a coarse-grained qualitative description of the underlying turbulent flow.
However the properties of these defects lines remain poorly understood and a clear connection with the transition to chaotic flow is still lacking.

In this work, we design an algorithm to detect nematic topological defects and carefully reconstruct the disclination lines in 3D active nematics.
This gives us access to the entire spectrum of defect lengths, from small loops to complex and tangled structures that span the whole system.
Depending on their topology, we distinguish a collection of regular defect loops, contractible onto a $3$-torus, coexisting with a background of wrapping defects.
Remarkably, these two classes of defects show different behaviour: as activity increases regular loops get smaller, while the length of the wrapping defects increases.
Furthermore, we study the morphology of regular defect loops and recognize a characteristic length scale, controlled by activity, that separates two different scaling regimes.

{\it The model}. The description of the nematic orientational order is given in terms of the symmetric and traceless nematic tensor field $Q_{ij}=q(n_in_j-\delta_{ij}/3)$.
$\underline{\bm{Q}}$ is invariant under local inversion of the nematic director $\bm{n}$, which represents the local orientation of the nematic field, $q$ being the magnitude of the local nematic tensor.
The nematic is embedded in an incompressible fluid of density $\rho$.
We make use of the well known \textit{Beris-Edwards} model for the coupled dynamics of the nematic field and the solvent~\cite{DeGennes_book,BerisEdwardsBook,thampi_turbulence}
\begin{align}
\label{eq:beris-nem}
    (\partial_t + \bm{u} \cdot \bm{\nabla}) \underline{\bm{Q}} - \underline{\bm{S}} &= \Gamma \underline{\bm{H}} \rm{,} \\
\label{eq:beris-fluid}
    (\partial_t + \bm{u} \cdot \bm{\nabla}) \bm{u} &= \frac{1}{\rho} \bm{\nabla} \cdot \underline{\bm{\Pi}}^{\rm{T}} \rm{.}
\end{align}
The nematic field is coupled to the flow, and its relaxation is governed by the Landau-de Gennes free energy with elastic coefficient $K$.
Activity induces local active stresses, $\underline{\bm{\sigma}}^{\rm{act}}=-\alpha \underline{\bm{Q}}$, with magnitude $\alpha$ (see Supplementary Material (SM)~\cite{sm} for a detailed description of the model).
We integrate numerically Eq.~(\ref{eq:beris-nem}) and ~(\ref{eq:beris-fluid}) in a 3D periodic cubic domain using a hybrid lattice Boltzmann (LB) finite-difference method~\cite{marenduzzoLB,cecilia_channel}.
The linear size of the simulation box is set to $L=192$, and a range of box sizes, from $L=96$ up to $L=768$, is used to perform finite-size analysis.
All lengths are given in lattice units.

The elastic and bulk free energy coefficients are chosen such that $\alpha$ swipes over two orders of magnitude, from $10^{-4}$ to $3 \times 10^{-2}$, still having the stationary state in the nematic phase.
Accordingly, the characteristic scale associated to the active stress, $l_{\rm{a}}=\sqrt{K/\alpha}$, ranges  from $l_{\rm{a}} \sim 10^{-1}$ to $l_{\rm{a}} \sim 10$. We consider extensile active stresses and flow aligning nematics, but quantitatively similar results hold for contractile activity and flow tumbling nematics, as shown in the SM~\cite{sm}.

{\it Defects' tracking.} In order to identify topological defects as singular points in the nematic orientational order, we extend the general idea introduced in~\cite{goldenfeld_defects}.
For uniaxial nematics in 3D, defects are organised into disclination lines.
Winding around a closed contour that surrounds the defect in its vicinity, concatenated with the defect line, the director field defines a map from the path to the projective plane $\mathbb{RP}^2$, which is the unit sphere with antipodal points identified.
After continuous deformation over the sphere, the maps associated to disclinations are the ones that connect two antipodal points~\cite{mermin_defects,alexander_defects}.
Traversing a contour that does not concatenate any defects returns a map that is contractible to a single point.
On this basis, we build a defects' tracking algorithm that is parameter-free and returns a set of defect points in space.
Once all defect points have been identified, they are clustered according to a prescribed criterion of proximity between points on the lattice~\cite{sm}.
Within the clusters, all defects and the corresponding neighbours network define a connected undirected graph.
Notably, each graph built according to this protocol has an even degree.
Under these circumstances, Euler's theorem ensures that a map exists between the cluster of defect points and a closed path that visits all defects exactly once~\cite{graphs}.
All clusters can therefore be described as closed polygonal curves.

\begin{figure}[t!!!!]
  \includegraphics[width=\columnwidth]{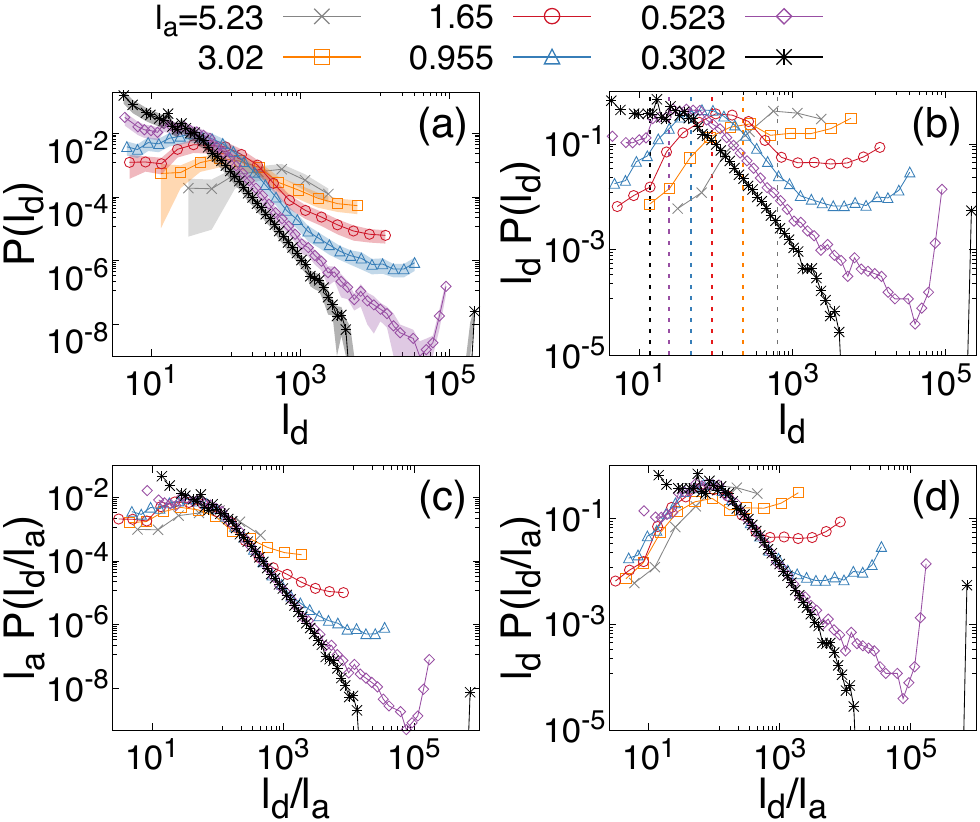}
  \caption{(a) Probability density function of the length of defect lines, for different values of the active length $l_{\rm a}$ (reported in the legend) with corresponding errors (shaded area).
  (b) Volume occupation for all defect sizes, for the same values of activities.
  The vertical dashed lines indicate the location of the average defect length, $\langle \l_{\rm{d}} \rangle$, as computed from the distributions in panel (a).
  Panels (c) and (d) show that the distributions in the first row, with the defect size, collapse onto a universal curve, when rescaled by the active length $l_{\rm{a}}$.
  }
\label{fig:pdfs}
\end{figure}

{\it Results}. 
Fig.~\ref{fig:pdfs} (a) displays the probability density function (pdf) $P(l_{\rm{d}})$ of the defect lines length, measured as the sum of the length of all edges in the corresponding polygonal curve, for different values of $l_{\rm{a}}$.
All cluster size distributions are peaked around a preferred size, $\langle l_{\rm{d}} \rangle$, which decreases with activity. Hence, activity selects the average length of the disclination lines. Fig.~\ref{fig:pdfs} (b) shows the distribution $l_{\rm{d}} P(l_{\rm{d}})$, which is peaked at the same value $\langle l_{\rm{d}} \rangle (l_{\rm{a}})$, meaning that the latter is also the size of defects that statistically occupy most of the volume in the system, while Figs.~\ref{fig:pdfs} (c)-(d) show that the defect lines length pdf scale with the active length $l_a$.

\begin{figure}[t!!]
  \includegraphics[width=\columnwidth]{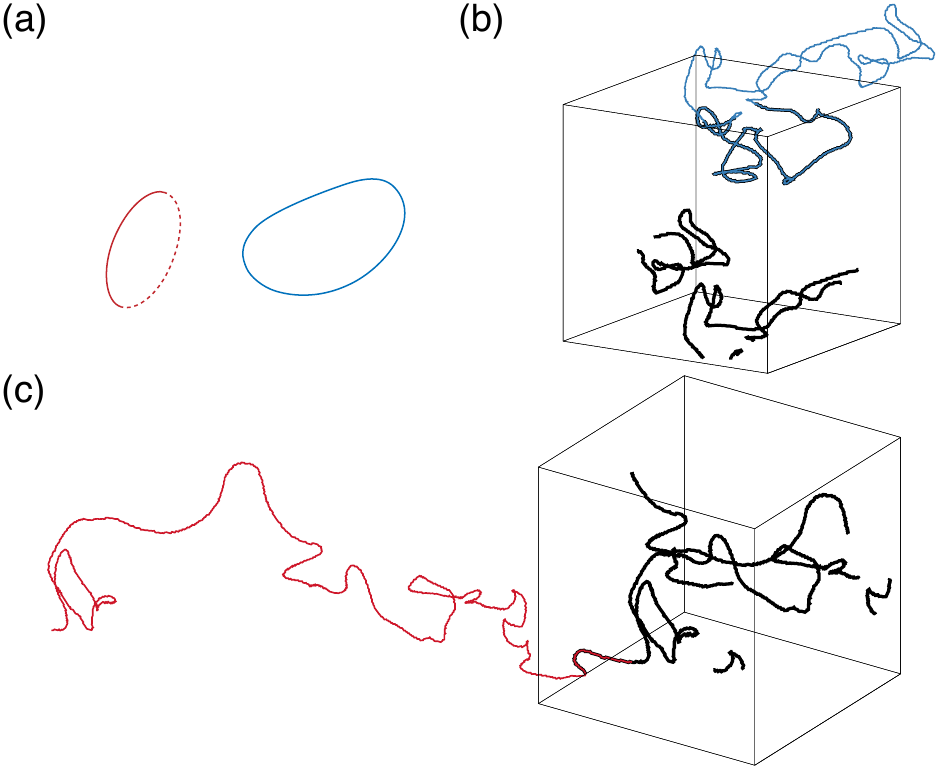}
  \caption{(a) Illustrative examples of one regular defect (blue) and one wrapping (red), respectively contractible and non contractible onto a $2$-torus, corresponding to a 2D periodic system. (b,c) Two disclination lines for active length $l_{\rm{a}}\simeq 3$. The thick black line corresponds to the disclination line tracked, while coloured lines show the same defect, unwrapped with respect to the periodic boundaries. In panel (a) the blue line corresponds to a regular loop, while in panel (b) the red line corresponds to a wrapping one. Note that only one defect line, out of the large number present, is shown for each configuration.
  }
\label{fig:snaps}
\end{figure}

In turn, Figs~\ref{fig:pdfs}(a) and \ref{fig:pdfs}(b) show that the system is populated by a wide distribution of defect lines, characterised by wide tail of the pdfs, whose length largely exceeds the system size, and whose excitation is allowed due to the system periodicity.
Under periodic boundary conditions (PBC), a defect line lives on a $3$-torus and can either close connecting to its initial point or to any other periodic images of the initial point, defining two different topological classes, as illustrated in Fig.~\ref{fig:snaps} (a) for a $2$-torus.
In the former case, referred to as {\it regular} defect, the contour can be contracted continuously to a point, while in the latter, referred to as {\it wrapping}, this is not possible.
Fig.~\ref{fig:snaps} (b) and (c) show an example of a regular and wrapping defect, respectively.

These two classes of defects contribute in a qualitatively different way to the defect tangle. Fig.~\ref{fig:pdfs} (c) shows that the defect pdf scales as $l_{\rm{a}}P(l_{\rm{d}}/l_{\rm{a}})$, except for the pfds' tails.
The pdf of the occupied volume fulfils the same scaling, as shown in Fig.~\ref{fig:pdfs} (d).

\begin{figure}[t!!!!]
  \includegraphics[width=\columnwidth]{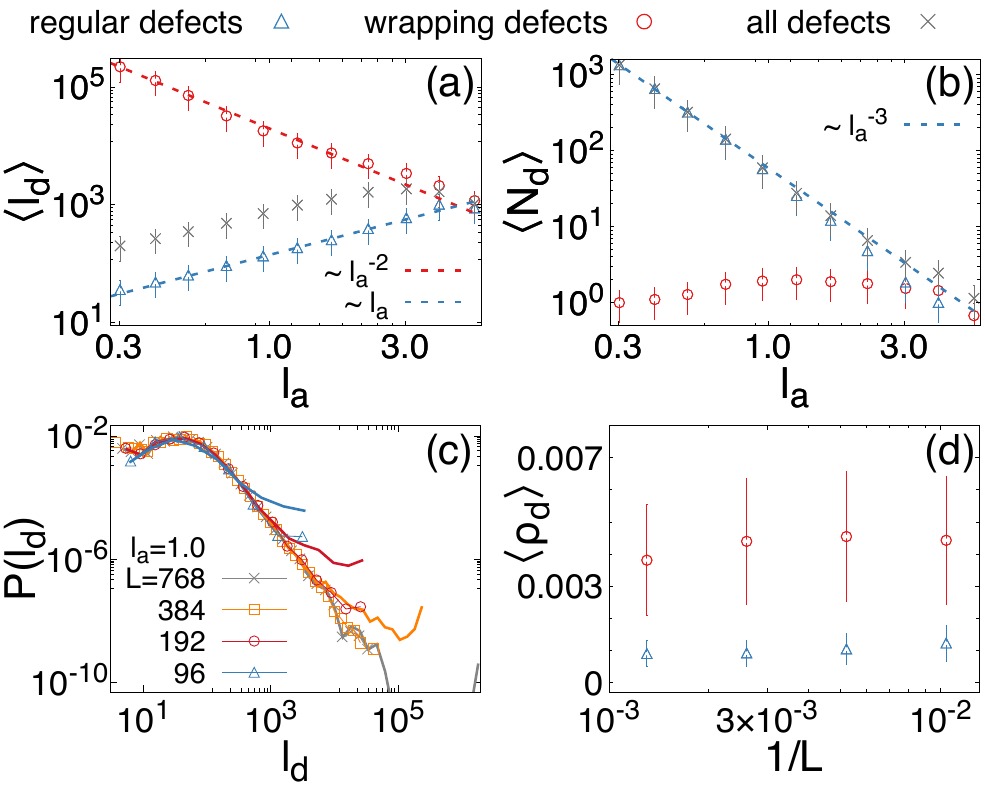}
  \caption{
  (a) Average defect length, corresponding to the first moment of the length distribution of Fig.~\ref{fig:pdfs}, as a function of the active length (grey crosses).
  The same quantity is reported, separately, for wrapping defects (red circles) and regular defects (blue triangles), with corresponding error bars.
  The corresponding scaling behaviour is also shown with dashed lines.
  (b) Average number of defect lines per configuration with the corresponding scaling behaviour.
  (c) Probability density function of defects length at $l_{\rm{a}}=1$, for different system sizes, indicated in the legend.
  The continuous lines represent the distribution for all defects, while symbols represent the distribution for regular defects only.
  (d) Volume fraction occupied by wrapping and regular defects, as a function of the system size, for $l_{\rm{a}}=1$.}
\label{fig:wrapping_def}
\end{figure}

Moreover, Figs.~\ref{fig:wrapping_def} (a,b) display the single defect average length, and average number of defect lines, respectively, as a function of $l_{\rm{a}}$, for wrapping defects (red), for regular defects (blue), and for all defects together (grey). The average linear size of regular defects scales as $\langle l_{\rm{d}} \rangle \sim l_{\rm{a}}$, and the average number of lines scale as $\langle N_{\rm{d}} \rangle \sim l_{\rm {a}}^{-3}$ while wrapping defects' average size grows as activity increases.
In particular, we verify that $\langle l_{\rm{d}} \rangle \sim l_{\rm{a}}^{-2}$, while the average number of these defects is always of order $\langle N_{\rm{d}} \rangle \sim 1$, independent on activity.

Note that, for $l_{\rm{a}}\gtrsim3$, the average number of defects with the two topologies is approximately the same, consistent with the fact that the system is close to the transition from chaotic to laminar flow.
Larger active lengths are no longer able to excite defects and the system relaxes to a long-range ordered state.
This behaviour of $N_{\rm{d}}$ results in a crossover from $\langle l_{\rm{d}} \rangle \sim l_a$ to $\langle l_{\rm{d}} \rangle \sim l_a^{-2}$ for the average length of all defects in the regime of large active length, as evident in the behaviour of the grey points in Fig.~\ref{fig:wrapping_def} (a).
Close to the transition to the chaotic state the average length for the two classes of defects is comparable, meaning that the easiest defects to sustain are system-size long line defects, either regular or wrapping.
As activity increases, the characteristic sizes of the two classes move away from each other, such that smaller regular defects proliferate and wrapping defects get longer.

Figs.~\ref{fig:wrapping_def} (c,d) display the impact of system size on both classes of defects. Panel (c) displays $P(l_{\rm{d}})$ for the same active length $l_{\rm{a}}=1$ and different system sizes, ranging from $L=96$ to $L=768$. The symbols represent the distribution of regular defects, while the continuous lines represent the distribution for all defects.
The shape of the pdf for regular defects does not change with $L$ and the peak is located always at the same value of $l_{\rm{d}}$.
However, larger system sizes allow to explore a wider range in the defects length, as is clear from the power law tail, and the average size of the wrapping defects is displaced to larger lengths as the system size increases.
In fact, the total length of a single wrapping defect, although it can exceed the system size with no a priori limits, is bounded from below: a defect shorter that the system size cannot wrap.
Even though the probability of finding wrapping defects of a given length $l_{\rm{d}}$ decreases upon increasing the system size, it does not vanish, as $\langle N_{\rm{d}} \rangle \sim 1$ up to the largest system sizes considered.
This behaviour is further clarified in Fig.~\ref{fig:wrapping_def} (d), which displays the overall volume fraction occupied by regular and wrapping defects as a function of the system size, defined as the ratio between the total length of the defects $L_{\rm d}$ and the system volume, $\rho_{\rm d} \sim L_{\rm d}/L^3$.
For all sizes considered and $\l_{\rm a} \simeq 1$, the contribution to the volume fraction from the wrapping defects is always larger than the one coming from the regular defects, and the two are approximately constant, ruling out finite size effects.
Hence, in the regime where defects proliferate, wrapping defects are always present and they can predominate in terms of total volume occupied.
The number of wrapping defects is small, suggesting that they behave as a topologically-preserved condensate that coexists with the broad distribution of regular defects.
We stress that wrapping disclinations are present and statistically relevant even though the topology of defects lines is not fixed and can turn from non-contractible to contractible and vice versa through local recombination of the tangle (see the movie in SM~\cite{sm}).

\begin{figure}[t!]
  \includegraphics[width=\columnwidth]{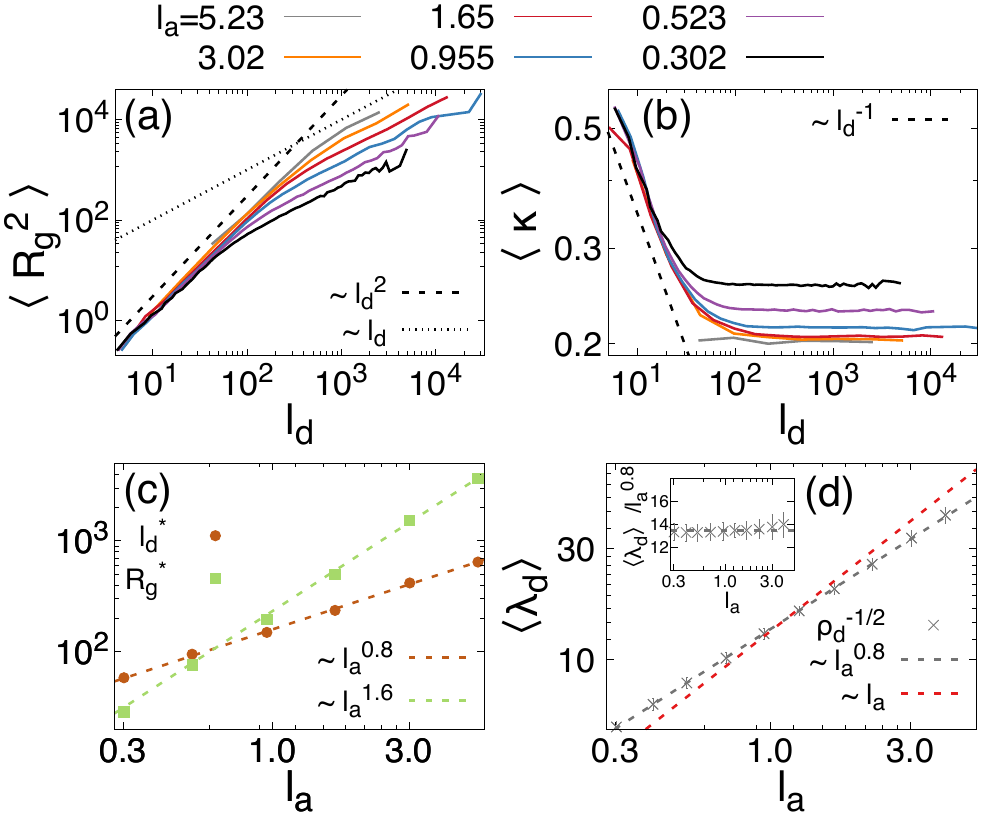}
  \caption{(a) Squared radius of gyration, ${R_{\rm g}^2=\sum_{i,j}{|\bm{r}_i-\bm{r}_j|^2}/2N^2}$, as a function of the defect length, for different values of activity.
  (b) Average curvature of defect lines as a function of $l_{\rm{d}}$.
  (c) Defect length and radius of gyration at the crossover between the two scaling regimes of panel (a), as a function of activity, with the corresponding scaling behaviour shown with dashed lines.
  (d) Average value of the estimated defects' separation, as a function of $l_a$, represented by the grey crosses, with the corresponding error. A power law $l_{\rm a}^{0.8}$ is shown with a dashed grey line, while $\sim l_{\rm{a}}$ is shown with a dashed red line for comparison. The scaling behaviour $\lambda_{\rm d}\sim l_{\rm a}^{0.8}$ is the best fit, as shown in the inset.}
\label{fig:radius_of_gyration}
\end{figure}

Activity strongly affects the defects' local curvature and their overall morphology.
Fig.~\ref{fig:radius_of_gyration} (a) displays the average squared radius of gyration, $\langle R_{{\rm g}}^2 \rangle$, of defect lines as a function of the their length, $l_{{\rm d}}$, for different values of $l_a$. 
For all activities considered, we clearly observe two separate length regimes, correlated to two different exponents in the power law scaling $R_{\rm{g}}^2\sim {l_{\rm{d}}}^{2/d}$, with $d$ the fractal dimension of the defect structures.
Small defects are characterised by $d=1$, which means that they are arranged into closed loops with a persistence length of the same order of the defect length.
From direct observation we conclude that these objects are rather planar, rigidly moving, small defect rings of regular shape.
Large defects show $d=2$, which, on the contrary, is associated to buckled structures with no preferred spatial direction.
In particular, the value of the exponent establishes that, independent of activity, in the limit of very large sizes, disclination lines are arranged as closed random walks.

Fig.~\ref{fig:radius_of_gyration} (b) displays the defects' average local curvature, $\langle \kappa \rangle$, as a function of $l_{{\rm d}}$.
For very small defects $\langle \kappa \rangle$ decreases, approaching the limiting power law $\langle \kappa \rangle \sim \l_{\rm{d}}^{-1}$, characteristic of planar shapes.
At larger sizes $\langle \kappa \rangle$ reaches a plateau, corresponding to constant curvature, tangled structures.
The value of the plateau depends on activity and, consistently with the results of $\langle R_{{\rm g}}^2 \rangle$, it increases with activity.
For the largest active length $l_{\rm{a}}=5.23$, we do not observe the first branch of the curve at smaller defects length, compatibly with the fact that, close to the transition to the chaotic regime, defects are long stretched lines of length comparable with the system size.
The crossover scale $l_{\rm{d}}^*$ between the two different structures scales with activity, $l_{\rm{d}}^*\sim {l_{\rm{a}}}^{\gamma}$  with $\gamma \simeq 0.8$, as shown in Fig.~\ref{fig:radius_of_gyration} (c).

In light of the above characterisation of the defects' morphology, we now look at the way the whole tangle is spatially arranged.
We measure the total length $L_{\rm{d}}$ of the defects tangle and the corresponding defect density $\rho_{\rm d}$, and we compute the average separation between defects as $\lambda_{\rm d} \sim \rho_{\rm d}^{-d/2}$.
Assuming that $\lambda_{\rm d} \sim l_{\rm a}$, as proposed in previous works~\cite{marchetti_actnem_corr,krajnik_spectra}, we expect $\rho_{\rm d}^{-1/2} \sim l_{\rm a}$ for small linear defects of $d=1$ and $\rho_{\rm d}^{-1/2} \sim l_{\rm a}^{1/2}$ for large buckled ones of fractal dimension $d=2$.
Comparing the location of the crossover length with the defects' length pdfs (see Fig.~\ref{fig:pdfs}), most defects populate the region around the crossover length.
No one of the two asymptotic morphologies dominate the defects tangle and we find an intermediate value of the exponent, $\rho_{\rm d}^{-1/2} \sim l_{\rm a}^{0.8}$.

{\it Conclusions}. 
As activity drives chaotic flows in nematic liquid crystals disclination lines proliferate. We have thoroughly characterized the nature of these defects and shown their morphological complexity. In a 3D periodic system defects belong to two topologically different classes, regular and wrapping disclination lines. While the number of regular defect lines increases with activity, the number of wrapping defects is of order one and can be regarded as a topologically preserved condensate.

The average length of the wrapping defects is proportional to the system size and increases upon decreasing the characteristic length scale, $l_{\rm a}$, while regular disclination loops get smaller and their length depends linearly on $l_{\rm a}$. The morphology and spatial organization of the regular defects strongly depends on their length. Small defect rings of planar shape and finite self-propelling speed coexist with long buckled defects that show the typical scaling exponent of random walks.  

Acknowledging this richness in defect morphology allows us to better rationalize the global properties of the defect tangle.
Specifically, the fact that the majority of defects belong to the crossover region between these two asymptotic classes, impacts the observed scalings, such as the dependence of the defect average separation length with activity. This opens new questions about the dynamics of disclination lines in relation to the heterogeneity in morphology observed and their topology in a periodic 3D system.

\noindent
{\it Acknowledgments} 
P.D. acknowledges support from the Juan de la Cierva-Formaci\'on program.
C.R. acknowledges funding from the European Union’s Horizon 2020 research and innovation program under the Marie Sklodowska-Curie Grant Agreement No. 754462.
I.P. acknowledges support from Ministerio de Ciencia, Innovaci\'on y Universidades MCIU/AEI/FEDER for financial support under grant agreement PID2021-126570NB-100 AEI/FEDER-EU, from Generalitat de Catalunya under Program Icrea Acad\`emia and project 2021SGR-673.
This work was supported by grants from the Swiss National Supercomputing Centre (CSCS), project IDs s1079 and s1194, and MareNostrum Supercomputer at Barcelona Supercomputing Center (BSC).

\bibliographystyle{apsrev4-1}
\bibliography{refs}

\clearpage
\widetext
\begin{center}
\textbf{\large Supplemental Material for \\
``Coexistence of defect morphologies in three dimensional active nematics''}

\date{\today}
\end{center}

\setcounter{equation}{0}
\setcounter{figure}{0}
\setcounter{table}{0}
\setcounter{page}{1}
\makeatletter
\renewcommand{\theequation}{S\arabic{equation}}
\renewcommand{\thesection}{S\arabic{section}}
\renewcommand{\thefigure}{S\arabic{figure}}

\section{Model and numerical simulations}

\begin{figure*}[bbb!]
  \includegraphics[width=\textwidth]{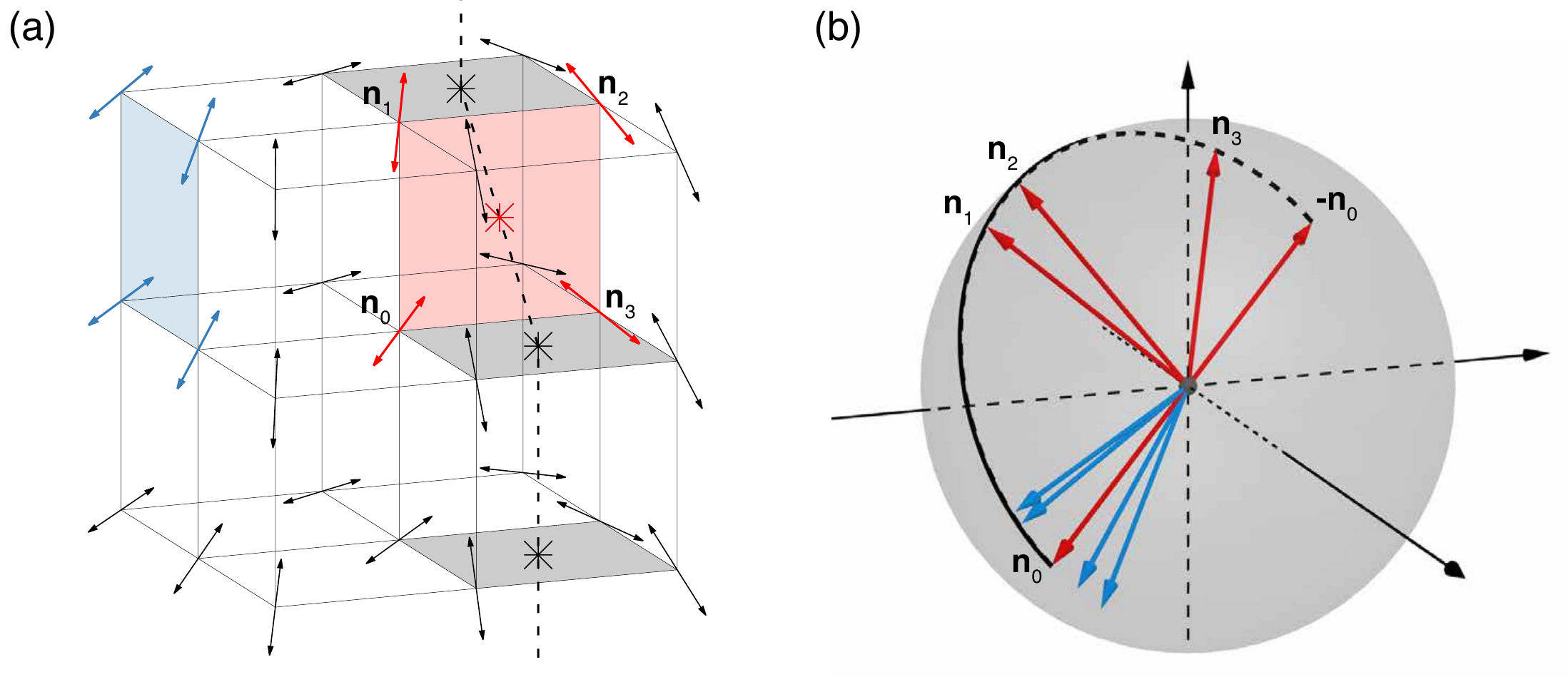}
  \caption{(a) A $3\times 3$ vertices portion of the system with the corresponding director field. Arrows are double-headed to represent the nematic symmetry.
  The face highlighted in blue colour is in the ordered phase, while the one highlighted in red corresponds to a defect, identified by the a red asterisk.
  Black faces also correspond to topological defect points, shown as black asterisks.
  The black dashed line is a disclination line joining the four point defects identified in the picture.
  (b) Mapping onto a unit sphere of the nematic field around the faces highlighted in blue and red in panel (a).
  The black line follows the path of the red arrows onto the spherical surface.
  }
\label{fig:one_voxel}
\end{figure*}

We write the equation of motion used for the evolution of a nematic liquid crystal embedded in a fluid in the presence of local active stress, and provide additional details about the model and the set of parameters adopted.
\begin{align}
\label{eq:sm_beris-nem}
    (\partial_t + \bm{u} \cdot \bm{\nabla}) \underline{\bm{Q}} - \underline{\bm{S}} &= \Gamma \underline{\bm{H}} \rm{,} \\
\label{eq:sm_beris-fluid}
    (\partial_t + \bm{u} \cdot \bm{\nabla}) \bm{u} &= \frac{1}{\rho} \bm{\nabla} \cdot \underline{\bm{\Pi}}^{\rm{T}} \rm{.}
\end{align}
Here, the nematic field $\underline{\bm Q}$ is advected by the flow and coupled to the flow gradients through the \textit{co-rotational term}
\begin{equation}
\label{eq:sm_corotational}
    \begin{aligned}
    \underline{\bm{S}} &=
    (\xi\underline{\bm{E}}+\underline{\bm{\Omega}}) \cdot (\underline{\bm{Q}}+\underline{\bm{I}}/3) + (\underline{\bm{Q}}+\underline{\bm{I}}/3) \cdot (\xi\underline{\bm{E}}-\underline{\bm{\Omega}}) \\
    &- 2\xi(\underline{\bm{Q}}+\underline{\bm{I}}/3) \ {\rm{tr}}(\underline{\bm{Q}} \cdot \underline{\bm{\nabla} \bm{u}}) \rm {.}
    \end{aligned}
\end{equation}
The character of this coupling is set by the flow aligning parameter $\xi$, while $\underline{\bm{E}}$ and $\underline{\bm{\Omega}}$ are the strain rate and the vorticity tensor, respectively.
All the results in the main text are given for the choice $\xi=0.7$, which corresponds to a regime of \textit{flow aligning}, where the director field tends to align to an underlying shear flow.
For values $\xi<0.5$ the nematic director field tends to rotate with respect to a shear flow.
This regime is called \textit{flow tumbling}, and we report later in this Supplementary Material results for this different coupling regime.
The nematic relaxation is governed by the molecular field $\underline{\bm{H}}=-\delta \mathcal{F}/\delta \underline{\bm{Q}}+\underline{\bm{I}} \ {\rm{tr}}(\delta \mathcal{F}/\delta \underline{\bm{Q}})$, with Landau-de Gennes free energy
\begin{equation}
\label{eq:sm_landau-de_gennes}
    \mathcal{F} = \int {\rm{dV}} \biggl[ \frac{K}{2}(\bm{\nabla}\underline{\bm{Q}})^2 + \frac{A}{2} {\rm{tr}}(\underline{\bm{Q}}^2) + \frac{B}{3} {\rm{tr}}(\underline{\bm{Q}}^3) + \frac{C}{4}( {\rm{tr}}(\underline{\bm{Q}}^2))^2 \biggr] \rm{.}
\end{equation}
The stress tensor is given by $\underline{\bm{\Pi}}=-P\underline{\bm{I}} + 2\eta\underline{\bm{E}} + \underline{\bm{\sigma}} + \underline{\bm{\sigma}}^{\rm{act}}$, with
\begin{equation}
\label{eq:sm_sigma-pass}
    \begin{aligned}
    \underline{\bm{\sigma}} &= 2\xi(\underline{\bm{Q}}+\underline{\bm{I}}/3) \ {\rm{tr}}(\underline{\bm{Q}} \cdot \underline{\bm{H}}) - \xi \underline{\bm{H}} \cdot (\underline{\bm{Q}}+\underline{\bm{I}}/3) \\
    &- \xi(\underline{\bm{Q}}+\underline{\bm{I}}/3) \cdot \underline{\bm{H}} - \bm{\nabla}\underline{\bm{Q}} \colon \frac{\delta \mathcal{F}}{\delta \bm{\nabla}\underline{\bm{Q}}} \\
    &+ \underline{\bm{Q}} \cdot \underline{\bm{H}} - \underline{\bm{H}} \cdot \underline{\bm{Q}} \ \rm{,}
    \end{aligned}
\end{equation}
and $\underline{\bm{\sigma}}^{\rm{act}}=-\alpha \underline{\bm{Q}}$.
The density and viscosity of the fluid are set to $\rho=1.0$ and $\eta=0.167$, respectively.
For the nematic liquid crystal, we set $K=0.002738$, $A=-0.2K$, $C=20.0K$, $B=-C$, $\Gamma=8.28$.
All parameters values are given in lattice units. Times are expressed in terms of the characteristic relaxation time $\tau=L^2/(K\Gamma)$, with $L$ the linear size of the system.
This choice of the parameters returns $q_0\simeq0.5$ in the ordered phase, and it allows us to keep a constant characteristic defect core size $l_{\rm{c}}\simeq1$~\cite{cecilia_channel}.

All measures are performed over $5$ independent simulations for each set of parameters used.
Each simulation runs over $10^6$ steps and configurations are sampled every $10^4$ steps.

\section{The defect tracking algorithm}

Topological defects in 3D nematics are named \textit{disclinations}, as they correspond to singularity in the orientational nematic field.
In order to identify singular points of the nematic orientation field, we examine all square contours corresponding to the faces of each 3D voxel in the lattice and establish if the face encloses a disclination, based on the construction depicted in Fig.~\ref{fig:one_voxel}.
While circuiting in an ordered way around the four vertices of the face, we construct a map of the local director field onto a unit sphere, as shown in Fig.~\ref{fig:one_voxel} (b).
As we deal with nematic symmetry, all opposite points on the sphere are identified.
Therefore, once we move to a new vertex, we choose, between the two equivalent (opposite) orientations of the director, the one that is closer to the previous one.
Using this procedure, the last step from the forth vertex back to the initial one returns two possible outcomes: either we go back to the initial point on the sphere (see the set of blue arrows in Fig.~\ref{fig:one_voxel} (b)), or we reach its antipodal (red arrows and the black path in Fig.~\ref{fig:one_voxel} (b)).
The latter case corresponds to the presence of a topological defect, which we place by convention in the center of the square, as indicated by the red asterisk in Fig.~\ref{fig:one_voxel} (a).

\begin{figure}
\includegraphics[width=\textwidth]{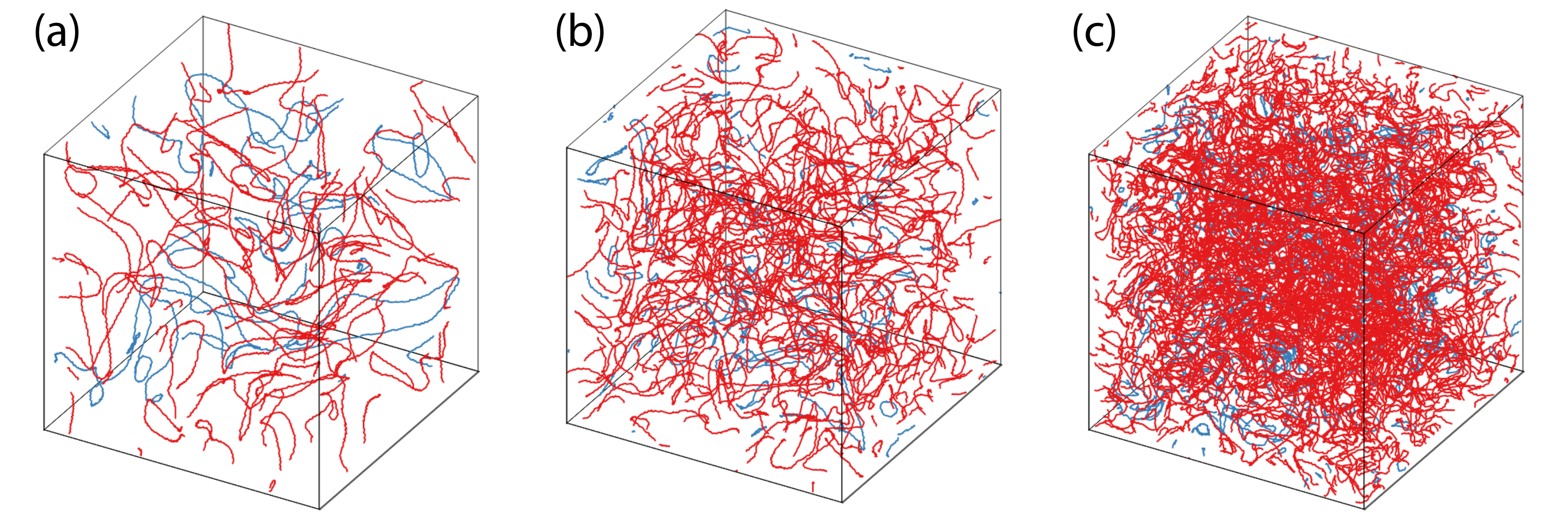}
\caption{Defect lines, as constructed with the method described in text for active length (a) $l_{\rm a}=1.65$, (b) $l_{\rm a}=0.955$, (c) $l_{\rm a}=0.523$.
Regular defects are shown in blue, while wrapping defects are shown in red.
}
\label{fig:confs}
\end{figure}

Once defect points have been found and a neighbourhood relationship has been assigned, we can group them into connected clusters.
In particular, we consider as neighbours two defect points if they belong to two adjacent faces or two opposite faces of the same voxel.
As described in the main text, we note that all clusters have even degree, which means that all nodes are associated to an even number of edges.
We can thus look for a closed path that visits all defect points in a cluster, travelling along each edge only once.
We identify such path by making use of the so-called \textit{Hierholzer's algorithm}~\cite{euler_algorithm,euler_book}.
As shown in Fig.~\ref{fig:confs}, this procedure returns a complete set of defect lines, that perfectly matches the defect pattern identified with regions of small scalar order parameter $q$.
See in the main text a detailed description of regular and wrapping defect lines, represented respectively in light blue and red in Fig.~\ref{fig:confs}.

\section{Flow tumbling regime and contractile activity}

\begin{figure}[ttt!!!!!]
\includegraphics[width=\textwidth]{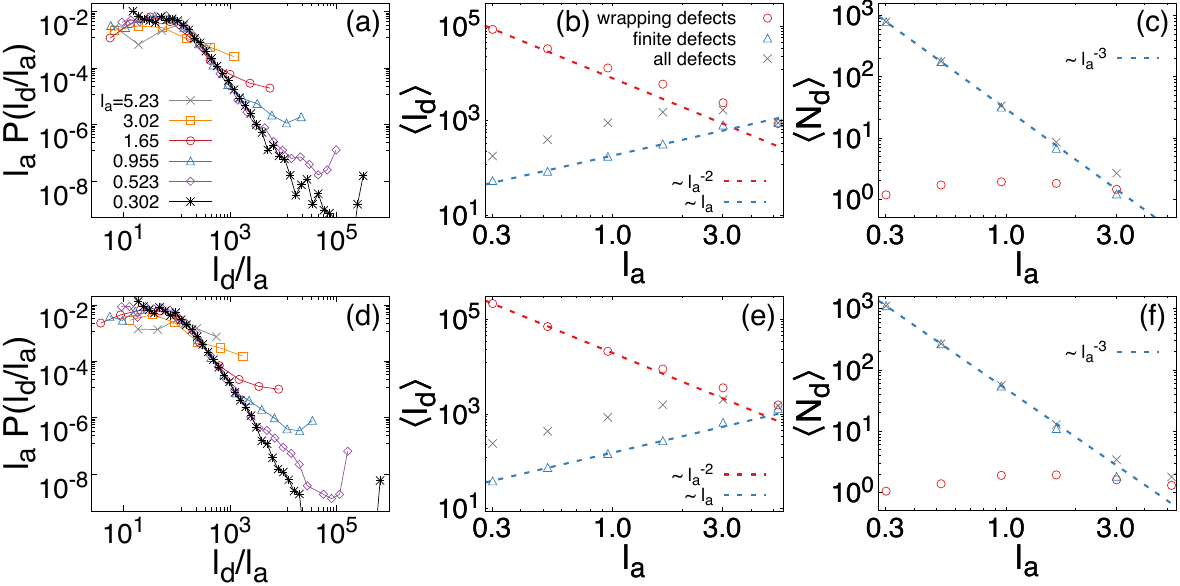}
\caption{(a,d) Probability density function of disclinations' length, as a function of $l_{\rm d}/l_{\rm a}$.
(b,e) Average length of the disclinations and (c,f) their average number, as a function of the active length.
The first row shows the results with contractile activity $\alpha<0$ and $\xi=0.7$, while $\alpha>0$ and $\xi=0.3$ in the second row.  
}
\label{fig:contractile}
\end{figure}

All the results reported in the main text refer to a regime of positive active parameter $\alpha$ and flow aligning coupling between the nematic field and underlying shear flow $\xi>0.5$.
There are however other choices available, and we will provide in this Section a set of results, corresponding separately to negative $\alpha$ and $\xi<0.5$.
It is well known that two main categories of swimmers exist, which are named \textit{pushers} and \textit{pullers}, depending whether they push or pull the flow along their swimming direction, respectively.
The sign of the parameter $\alpha$ in our model captures this different behavior.
In particular, positive values of $\alpha$ generate a puller-type local flow field and in this regime the active stress $-\alpha \underline{\bm Q}$ is called \textit{extensile}.
On the other hand, negative values of $\alpha$ correspond to puller swimmers and the activity is named \textit{contractile}.
We provide in Fig.~\ref{fig:contractile} (a-c) the scaled probability density function of the defect length, the average defect length and the average number of disclination lines, in the regime of contractile activity, for different values of active length $l_{\rm a}=\sqrt{K/|\alpha|}$.
Fig.~\ref{fig:contractile} (d-f) show the same quantities, in a regime of extensile activity, for the same set of values of $\l_{\rm a}$, but for $\xi=0.3$, in the flow tumbling regime.

Both the results with contractile activity and in the flow tumbling regime do not show notable differences from the ones with extensile activity and flow aligning, which are the focus of the analysis  in the main text.
The maximum in the probability density function of the defect length scales with active length, identifying a clear dependence of the length of the regular defects on activity,  is not modified either by the contractile nature of the activity or the flow-tumbling character of the coupling to the flow.
We also observe the presence of system-spanning wrapping defects, at all values of activity explored.
As shown in Fig~\ref{fig:contractile} (b,c) for contractile activity and (e,f) for flow tumbling, the average length of these percolating network of defects increases as $\alpha$ grows in magnitude, with a power law dependence on active length $\langle l_{\rm d} \rangle \sim l_{\rm a}^{-2}$.
The average number of percolating defects is always of order $\langle N_{\rm d} \rangle \sim 1$, independently on activity.

\section{Scaling crossover}

\begin{figure}[h!!!!!]
\includegraphics[width=\textwidth]{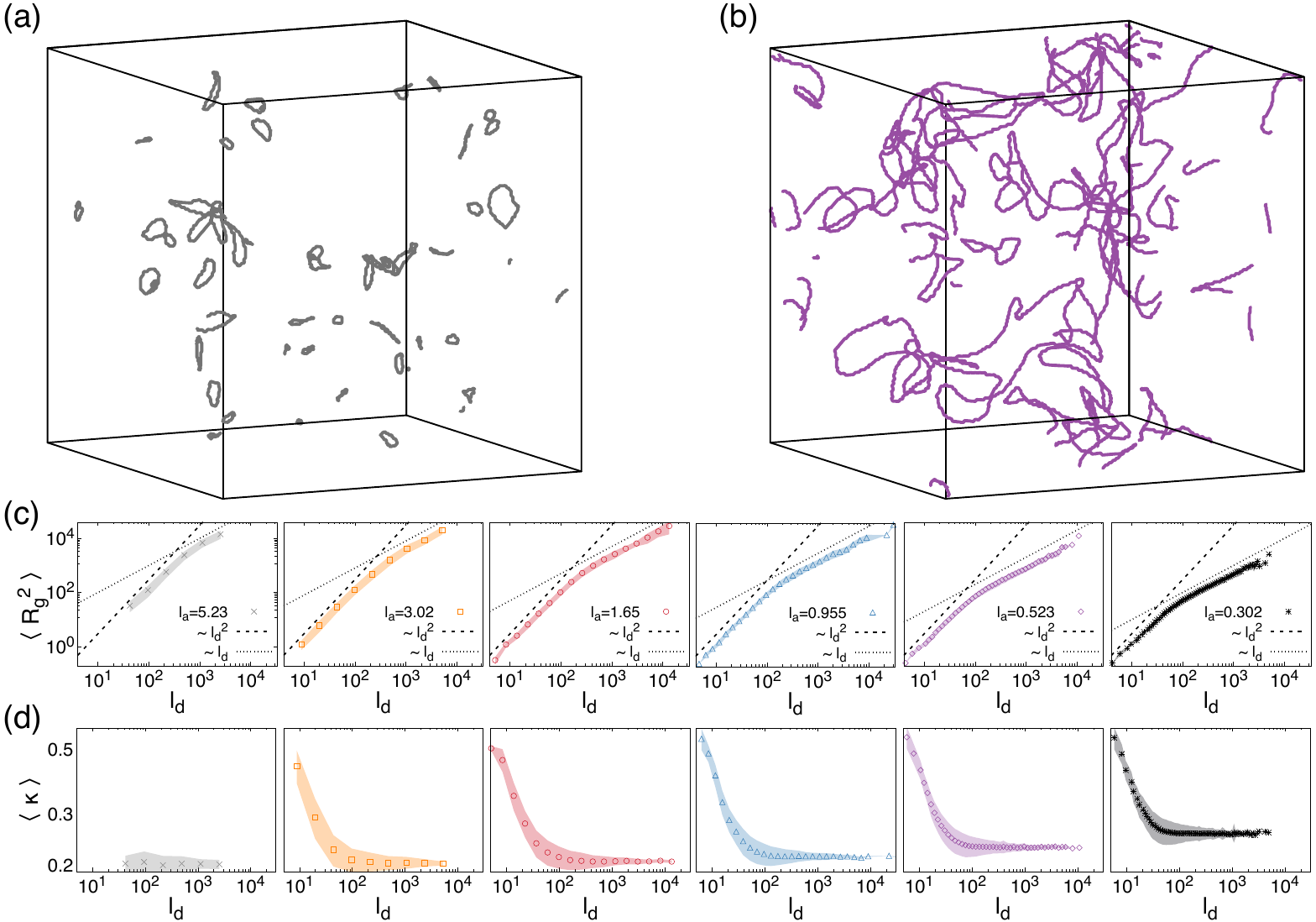}
\caption{(a) Disclination loops of length $l_{\rm d}<l_{d}^*$ and (b) $l_{\rm d}<l_{d}^*$.
(c) Average radius of gyration as a function of the defect length, for different values of activity, given in the legend. The same curves as in Fig. 4 are shown, but in separate panels and with the corresponding statistical uncertainty.
(d) Average local curvature, as a function of the defect length, for the same values of activity, as shown in Fig. 4, with corresponding errors.
}
\label{fig:crossover_length}
\end{figure}

We provide supplemental pictures concerning the crossover length between the two scaling regimes of defect lines, reported in Fig. 4 of the main text and therein commented.
Small defects are closed loops of approximately planar shape, and are coherently described bu a scaling exponent $R_{\rm g}^2 \sim l_{\rm d}^2$. See the typical shape of these defects in Fig.~\ref{fig:crossover_length} (a).
On the other hand, larger defects show a scaling exponent $R_{\rm g}^2 \sim l_{\rm d}$.
Therefore, their morphology is the one of a random walk in 3D, with a given curvature that depends on activity. Examples of these defects are given in Fig.~\ref{fig:crossover_length} (b).
The crossover length between these two groups of defects strongly depends on activity, according $l_{\rm d}^* \sim l_{\rm a}^{0.8}$ (see Fig. 4 in the main text).
We locate the crossing point with an iterative protocol.
We make a first guess of the point $l_{\rm{d}}^*$ and fit the two branches restricted to the range $l_{\rm{d}} < l_{\rm{d}}^*$ and $l_{\rm{d}} > l_{\rm{d}}^*$, respectively.
This returns a new estimate of the crossing point $l_{\rm{d}}^*$.
The same procedure is iterated until convergence.
Fig.~\ref{fig:crossover_length} (c,d) show the average radius of gyration and the average local curvature, respectively, as a function of the defect length, with corresponding uncertainty on the average.

\section{Supplemental movie}

The supplemental movie ``regular-wrapping\_defs\_alpha0.001.mov"~\cite{movies} show the typical dynamics of the disclination lines, as tracked with our algorithm.
The evolution starts from a nematically ordered state and active length $l_{\rm a}=1.65$.
Very quickly the systems starts to develop topological defects, that grow in number and length, until they reach a stationary global density with regular and wrapping lines.
The panel on the left shows the whole defect tangle with regular defects coloured in blue and wrapping defects coloured in red.
In panel on the right the same evolution is shown, with only regular defects visible.
Defect lines evolve in time and continuously recombine between each other, in a way that the length and shape of single lines are not conserved quantities.
The topology of defect is not protected and they can turn from regular to wrapping and vice versa with no a priori constraints.

\end{document}